\documentclass[twocolumn]{aastex631}

\usepackage{hyperref}
\hypersetup{
    colorlinks=true,
    linkcolor=blue,
    filecolor=magenta,      
    urlcolor=cyan,
}
\usepackage{xspace}
\usepackage{float}
\usepackage[version=4]{mhchem}

\usepackage[T1]{fontenc}
\newcommand{\ms}{\ensuremath{\mathrm{m}\,\mathrm{s}^{-1}}\xspace}
\newcommand{\kms}{\ensuremath{\mathrm{km}\,\mathrm{s}^{-1}}\xspace}

\renewcommand{\Re}{\ensuremath{R_{\oplus}}\xspace}

\newcommand{\Rsun}{\ensuremath{R_{\odot}}\xspace }
\newcommand{\Msun}{\ensuremath{M_{\odot}}\xspace}

\begin{document}
\
\title{Measuring the Obliquities of the TRAPPIST-1 Planets with MAROON-X}

\author[0000-0003-2404-2427]{Madison Brady}
\affiliation{Department of Astronomy \& Astrophysics, University of Chicago, Chicago, IL 60637, USA}

\author[0000-0003-4733-6532]{Jacob L.\ Bean}
\affiliation{Department of Astronomy \& Astrophysics, University of Chicago, Chicago, IL 60637, USA}

\author[0000-0003-4526-3747]{Andreas Seifahrt}
\affiliation{Department of Astronomy \& Astrophysics, University of Chicago, Chicago, IL 60637, USA}

\author[0000-0003-0534-6388]{David Kasper}
\affiliation{Department of Astronomy \& Astrophysics, University of Chicago, Chicago, IL 60637, USA}

\author[0000-0002-4671-2957]{Rafael Luque}
\affiliation{Department of Astronomy \& Astrophysics, University of Chicago, Chicago, IL 60637, USA}

\author[0000-0003-1242-5922]{Ansgar Reiners}
\affiliation{Institut f{\"u}r Astrophysik und Geophysik, Georg-August-Universit{\"a}t, Friedrich-Hund-Platz 1, D-37077 G{\"o}ttingen, Germany}

\author[0000-0001-5578-1498]{Bj{\"o}rn Benneke}
\affiliation{University of Montreal, Montreal, QC, H3T 1J4, Canada}

\author[0000-0001-7409-5688]{Gudmundur Stef\'ansson}
\affiliation{Department of Astrophysical Sciences, Princeton University, 4 Ivy Lane, Princeton, NJ 08540, USA}
\affiliation{NASA Sagan Fellow}

\author[0000-0002-4410-4712]{Julian St{\"u}rmer}
\affiliation{Landessternwarte, Zentrum f{\"u}r Astronomie der Universität Heidelberg, K{\"o}nigstuhl 12, D-69117 Heidelberg, Germany}

\begin{abstract}
A star's obliquity with respect to its planetary system can provide us with insight into the system's formation and evolution, as well as hinting at the presence of additional objects in the system.  However, M dwarfs, which are the most promising targets for atmospheric follow-up, are underrepresented in terms of obliquity characterization surveys due to the challenges associated with making precise measurements.  In this paper, we use the extreme-precision radial velocity spectrograph MAROON-X to measure the obliquity of the late M dwarf TRAPPIST-1.  With the Rossiter-McLaughlin effect, we measure a system obliquity of $-2\degr ^{+17\degr}_{-19\degr}$ and a stellar rotational velocity of 2.1 $\pm$ 0.3\,\kms.  We were unable to detect stellar surface differential rotation, and we found that a model in which all planets share the same obliquity was favored by our current data.  We were also unable to make a detection of the signatures of the planets using Doppler tomography, which is likely a result of the both the slow rotation of the star and the low SNR of the data.  Overall, TRAPPIST-1 appears to have a low obliquity, which could imply that the system has a low primordial obliquity.  It also appears to be a slow rotator, which is consistent with past characterizations of the system and estimates of the star's rotation period.  The MAROON-X data allow for a precise measurement of the stellar obliquity through the Rossiter-McLaughlin effect, highlighting the capabilities of MAROON-X and its ability to make high-precision RV measurements around late, dim stars.

\end{abstract}

\section{Introduction}
\label{sec:intro}

A star's obliquity ($\lambda$) is the angle between the stellar angular momentum and its planets' orbital angular momenta, and it gives us insight into the formation and evolution of its planetary system.  As an example, the solar system’s obliquity is $7\degr$ \citep{Beck05}.  This low but non-zero obliquity has been explained with a misaligned protoplanetary disk \citep{Wijnen17}, past gravitational encounters with other stars \citep{Cuello22}, asymmetric solar winds \citep{Spalding19}, and additional giant planets \citep{Bailey16}.  Furthermore, exoplanet systems have been discovered with enhanced obliquities in excess of $80\degr$ \cite[see, e.g.,][]{Albrecht12, Dalal19}, which could be explained by dynamical interactions with other planets or stars \citep[see][and the references therein]{Albrecht22, Louden21}.

If an enhanced obliquity hints at the presence of additional giant planets in the system, it could also tell us about the system's habitability, as distant giant planets have been shown in \cite{Clement22} to enhance the delivery of volatiles from beyond the system snowline to the star's habitable zone.  The delivery of water would enhance the chances of the interior planets having liquid water oceans.  However, this relationship is still poorly constrained, and giant planets may not be required for obliquity enhancements \citep{Louden21}.  Even if an observed planet’s obliquity is not enhanced by additional planets in the system, the obliquity can provide us with insights into the host star's magnetism early in its life \citep{Lai11}.  

The Rossiter-McLaughlin \citep[RM; ][]{Rossiter24, McLaughlin24} effect is commonly used to measure stellar obliquities of systems with transiting planets.  As a planet passes over its host's surface, it eclipses portions of the stellar surface that are redshifted or blueshifted due to the star's rotation.  This causes a slight perturbation in the star's line profiles that manifest as apparent radial velocity (RV) shifts.  These perturbations scan the planet's path over the star's surface and can tell us the angle between the star's rotation axis and the plane of the planet's orbit.   Of course, this method requires a transiting planet and sensitive RV measurements, so it is best suited to short-period planets orbiting bright stars with rich spectra and moderate rotational velocities. 

Currently, hot host stars ($T_{eff}>6000$\,K) are observed (primarily via the RM effect) to frequently host high-obliquity exoplanets, while cooler stars tend to host planets with very low obliquities.  These observations could be explained if the thicker convective envelopes of cool stars are substantially more effective at tidally re-aligning planetary systems with respect to the host star's rotation axis than the radiative envelopes of hot stars \citep{Winn10}.  M dwarfs, which have deep convective layers \citep[or, for $M_\star < 0.35\,\Msun$, are fully convective; see][]{Chabrier97} could in principle give us insight into the relationship between system obliquity and host star convection.  However, studying this mechanism is complicated by the fact that the tidal dissipation timescale is also a strong function of $a/R_\star$ \citep[see, e.g.,][]{Zahn77}, where small stars are expected to be substantially less effective at aligning their planetary orbits.  Longer-period planets around M dwarfs may thus be excellent probes into primordial misaligning mechanisms.  

However, $\lambda$ is difficult to measure for M dwarfs because they are small and dim and only rarely host large planets (which have stronger RM signals).  While most observed M dwarfs seem to have low obliquities \citep{Hirano2020, Hirano20b, Stefansson20}, at least two systems \citep[Gl436 and GJ3470;][]{Bourrier18, Stefansson22} appear to have excited obliquities consistent with polar orbits.  These obliquities could provide us with insight into the primordial obliquity distribution of these planets, tidal damping theory, or be signatures of external perturbers.  However, as many of these measurements have large errors, it is necessary to continue gathering and refining the obliquity measurements of nearby M dwarf systems.

TRAPPIST-1 is an interesting system for obliquity measurements, as it is one of the closest and smallest fully-convective systems with observed planetary transits.  TRAPPIST-1 is a very small ($R_\star = 0.12\,\Rsun$) and cool ($T_{\mathrm{eff}} = 2600$\,K) late M dwarf that hosts seven well-characterized planets (b, c, d, e, f, g, and h) with radii between 0.8 -- 1.1\,$R_\oplus$, of which three are in or near the habitable zone \citep{Gillon16, Gillon2017, Agol2021}.  The system is currently the target of more than one hundred hours of JWST time, which will give us substantial insight into the properties of these planets' atmospheres.  Constraints from photometry and transit timing variations show that the known planets fall on coplanar (or near-coplanar) orbits \citep{Agol2021}.  \citep{Hirano2020} performed RM measurements of the b, e, and f planets with the IRD spectrograph on Subaru \citep{Kotani18}.  They were unable to constrain the obliquities separately, but did find that the planets together constrained the overall system obliquity to $\lambda=1\degr \pm 28\degr$ with the RM effect.  They also claimed a measurement of $\lambda= 19\degr ^{+13\degr}_{-15\degr}$ with Doppler tomography, though they only detected the tomographic signal at the 1.9\% false alarm probability threshold for planet b and at much higher false alarm probabilities for e and f.  These measurements point towards a low system obliquity.  

This low obliquity is consistent with the hypothesis that stars with deep convective regions are efficiently capable of re-aligning their planets.  However, the TRAPPIST-1 planets are expected to have relatively long tidal damping timescales, which would allow them to retain any primordial obliquity excitation \citep[see Figure 22 from][]{Albrecht22}.  This indicates that either the tidal damping mechanism is more effective than previously thought or that the system formed at a zero obliquity.  However, an improvement in measurement precision is necessary to draw any concrete conclusions about the system and its dynamical history.

A substantial obliquity could be a hint that external massive companions may be present in the TRAPPIST-1 system.  External giant planets can strongly impact the presence and habitability of inner rocky planets \citep[see, e.g.,][]{Schlecker21, Vervoort22}.  While it might seem far-fetched to think that an additional giant planet could enhance the system obliquity without exciting the planets' mutual inclinations, \cite{Gratia2017} found that multiple giant planets can torque close-packed planetary systems without strongly impacting their mutual inclinations.  An external massive companion is believed to be responsible for the high-obliquity coplanar planets in the K2-290 system \citep{Hjorth21}.  In addition, \cite{Bryan2019} found that long-period Jupiters are common in Super-Earth systems (around a 40\% occurrence rate), and in fact seem to be more common in such systems than around random field stars.  Thus, large planets could help facilitate the formation of close-packed systems like TRAPPIST-1.  

However, TRAPPIST-1 is an extremely challenging star to study.  It is very dim, with a $V$-band magnitude of 18.8 \citep{Costa06}.  Combined with a rotational velocity on the order of $<3$\,\kms \citep[e.g.,][]{Hirano2020}, the expected RM amplitude of most of the TRAPPIST-1 planets falls between 5 -- 10\,\ms.  This, combined with short transit durations (40 -- 80\,minutes), makes it difficult to perform RV measurements which are simultaneously precise enough to measure the RM amplitude and have sufficient temporal resolution to capture the shape of the RM curve.  This necessitates the usage of a precise RV spectrograph on a large telescope, as well as multiple transit observations.

MAROON-X \citep{Seifahrt18} is a precise RV spectrograph ideally suited to performing such measurements of TRAPPIST-1. A ten-minute exposure from MAROON-X of TRAPPIST-1 has a RV error of around 5\,\ms, making a direct measurement of the RM effect feasible with this instrument.  In this paper, we discuss a newly-collected set of MAROON-X RVs measuring the RM effect of planets b, c, d, e, f, and g, and use these RVs to constrain the obliquities of the individual planets, as well as the obliquity of the seven-planet system as a whole.  This study additionally serves to show MAROON-X's capabilities as an instrument for the precise performance of the RM effect on dim, late-type stars.  In Section~\ref{sec:obs}, we describe MAROON-X and our observations.  In Section~\ref{sec:analysis}, we describe our efforts to perform a fit to the MAROON-X data with the RM effect and Doppler tomography.  In Section~\ref{sec:results}, we discuss our results and conclude.

\section{MAROON-X Observations}
\label{sec:obs}

\begin{figure*}
    \centering
    \includegraphics[width=\linewidth]{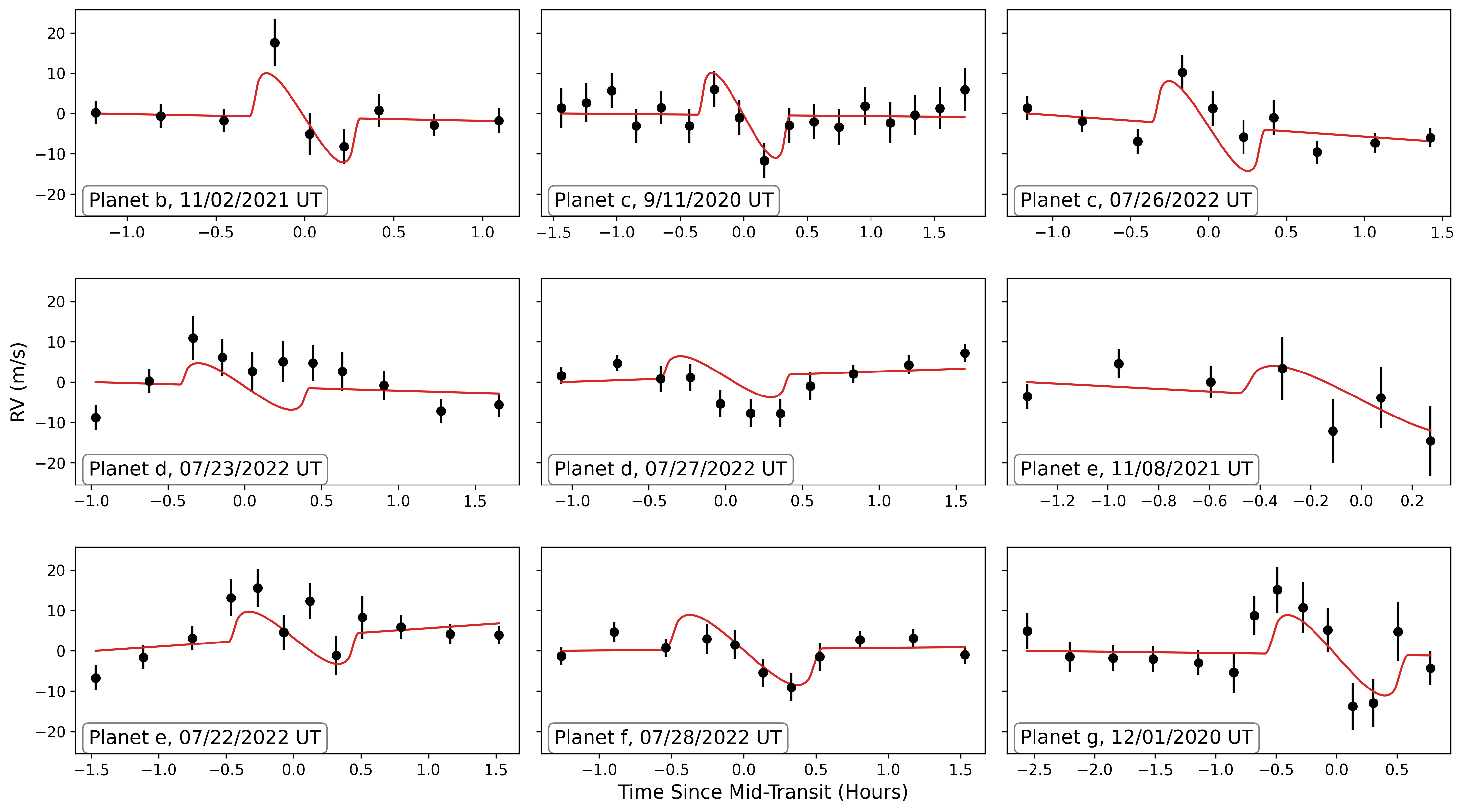}
    \caption{The observed RVs for the TRAPPIST-1 planets, with the best-fit RM effect model overplotted in red.  Each individual transit is plotted separately.  The x-axis is expressed in time since mid-transit, with the mid-transit time defined as that corresponding to the fit values from Section~\ref{ssec:rm}.}
    \label{fig:rm_rv}
\end{figure*}

MAROON-X is a stabilized, high-resolution echelle spectrograph designed primarily for measuring precise radial velocities \citep{seifahrt16,Seifahrt18,seifahrt20,seifahrt22}.  With a wavelength coverage of 500 -- 920\,nm, it is ideally suited for measuring RVs of M dwarfs.  The instrument possesses two optical-NIR channels with different wavelength coverage (500 -- 670\,nm and 650 -- 900\,nm).  Both channels are exposed simultaneously when observing a target.  However, TRAPPIST-1 is so cool and dim that the blue channel data had too low a signal-to-noise ratio (SNR) to be usable.  All of the data presented in this paper are from the redder channel exclusively, and are shown in Figure~\ref{fig:rm_rv} alongside a RM model described in Section~\ref{ssec:rm}.

To obtain the maximal precision on the TRAPPIST-1 obliquity in the shortest time span, we attempted to observe as many individual planet transits as possible.  We observed TRAPPIST-1 nine times between September 2020 and July 2022, overall observing eight full transits (the transit of TRAPPIST-1e on 8 November 2021 UT was aborted halfway through due to degrading weather conditions).  None of the observed transits overlapped with one another.  As TRAPPIST-1 is very dim, we only attempted observations in excellent weather conditions, with $<0.3$ magnitude loss from clouds and $<0.75"$ seeing at zenith.  These restrictive weather requirements were necessary in order to achieve the desired RV precision.  The individual observations are described in Table~\ref{tab:t1_obs}.  The majority of the transits were taken during July 2022, primarily due to unusually good weather conditions during that month.  We were able to observe transits of the six planets b, c, d, e, f, and g, only missing  planet h.  

The different planets offer different advantages when it comes to measuring the RM signal.  Planets like b and c have relatively large expected signal amplitudes due to their large radii, but tend to have short enough transits that it is difficult to capture the precise shape of the RM curve while simultaneously collecting sufficiently long exposures to beat down the noise.  Meanwhile, planets like g have long enough transits that phase coverage is less of an issue, but transit so rarely that it is difficult to collect multiple transits.  We attempted to observe planets with weaker expected signals or shorter transits multiple times in order to derive better constraints on their obliquities, but were not always able to do so due to the very strict weather requirements of these observations. 

\begin{table*}[]
\centering
\begin{tabular}{ccccccc}
\textbf{UT Date}       & \textbf{Planet} & \textbf{n$_{\mathrm{obs}}$ (20, 10\,min)} & \textbf{Airmass} & \textbf{$\sigma_{RV}$ (m$s^{-1}$) (20, 10\,min)}  & \textbf{K$_{\mathrm{RM}}$ (m$s^{-1}$)} & \textbf{$T_{\mathrm{dur}}$ (min)} \\ \hline
09/11/2020 10:29-13:40 & c               & 0, 17                                                                       & 1.11 -- 1.84     & --, 4.6                                                                                     & 10                                   & 42.03 $\pm$ 0.13         \\
12/01/2020 05:27-08:47 & g               & 6, 8                                                                        & 1.11 -- 2.14     & 3.7, 5.8                                                                                    & 10                                   & 68.24 $\pm$ 0.28         \\
11/02/2021 07:07-09:23 & b               & 5, 4                                                                        & 1.10 -- 1.40     & 2.9, 4.9                                                                                    & 10                                   & 36.06 $\pm$ 0.11         \\
11/08/2021 06:29-08:05 & e               & 3, 4                                                                        & 1.10 -- 1.22     & 3.6, 7.9                                                                                    & 7                                    & 55.76 $\pm$ 0.26         \\
07/22/2022 10:12-13:11 & e               & 6, 6                                                                        & 1.81 -- 1.11     & 2.8, 4.6                                                                                    & 7                                    & 55.76 $\pm$ 0.26         \\
07/23/2022 10:03-12:41 & d               & 5, 6                                                                        & 1.87 -- 1.13     & 3.1, 4.8                                                                                    & 5                                    & 48.87 $\pm$ 0.24         \\
07/26/2022 09:34-12:09 & c               & 6, 4                                                                        & 2.09 -- 1.16     & 2.7, 4.2                                                                                    & 10                                   & 42.03 $\pm$ 0.13         \\
07/27/2022 11:08-13:45 & d               & 5, 6                                                                        & 1.32 -- 1.11     & 2.2, 3.4                                                                                    & 5                                    & 48.87 $\pm$ 0.24         \\
07/28/2022 09:16-12:03 & f               & 6, 5                                                                        & 2.26 -- 1.16     & 2.3, 3.5                                                                                    & 9                                    & 62.85 $\pm$ 0.25        
\end{tabular}
\caption{Observed TRAPPIST-1 planet transits, along with the number of exposures, airmasses of observations, and average observed RV error.  We also include the expected transit duration $T_{\mathrm{dur}}$ from \cite{Agol2021} and the estimated RM signal amplitude K$_{\mathrm{RM}}$, assuming a zero obliquity and stellar rotational period of 3.3 days.}
\label{tab:t1_obs}
\end{table*}

In all except the very first observation (TRAPPIST-1c on September 11, 2020), we performed 20-minute exposures while the planet was not transiting in order to reduce the scatter on our observed baseline and 10-minute exposures during the planet transit to reduce the velocity smearing of the RM signal.  The transit times used for the purpose of scheduling observations were taken from \cite{Agol2021}, which provides forecasts for TRAPPIST-1 transit times up through 2023, with estimated errors between half a minute and five minutes.  We observed the target for an hour before and after the transit to get an adequate baseline for subtracting out the orbital motion of the star and any activity signals, so each transit observation took around three hours to complete.

The MAROON-X data were reduced using a custom \texttt{Python3} pipeline developed using tools for the CRIRES instrument \citep{Bean10} and RVs were calculated with a version of \texttt{serval} \citep{Zechmeister20} modified to work with MAROON-X data.  \texttt{serval} calculates RVs by stacking the observed target spectra to form a template, which is then RV shifted and compared to each individual observation using least-squares fitting.  This method tends to produce more precise RVs for M dwarfs than the familiar cross-correlation function (CCF) method, which relies more heavily upon the usage of binary line lists and continuum fitting, both of which are harder to accomplish with M dwarfs, which have many blended lines and often lack an obvious continuum in the visible and NIR regime \citep{Anglada-Escude12}.  Telluric lines from the Earth's atmosphere are masked for the purpose of producing our RVs.  Our spectral-wavelength calibration comes from a combination of a Fabry-Perot Etalon-fed fiber simultaneously recorded with all science frames and a series of calibration exposures including the usage of a Thorium-Argon Hallow Cathode Lamp, as described in \cite{seifahrt22}. 

\begin{figure}
    \centering
    \includegraphics[width=1\linewidth]{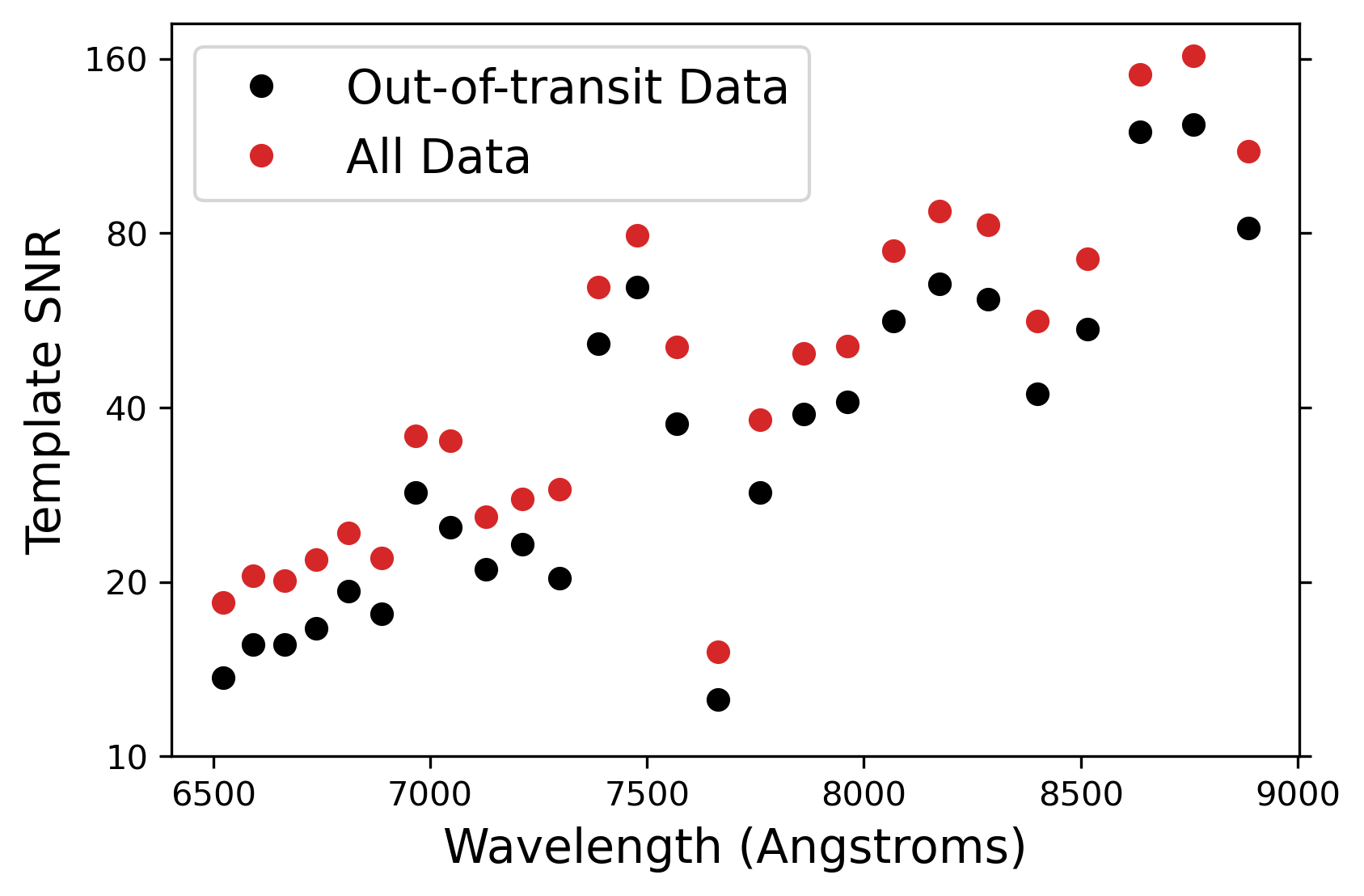}
    \caption{The peak SNR of the each template spectral order versus wavelength, in angstroms.  The template including only out-of-transit data is shown in black, while the template including all TRAPPIST-1 data is shown in red.  Only data from the redder channel of MAROON-X are included, as the blue channel data have very low signals.}
    \label{fig:template_SNR}
\end{figure}

We used \texttt{serval} to create a high SNR TRAPPIST-1 template spectrum (Figure~\ref{fig:template_SNR}) with out-of-transit data, and then used this template to measure the RM signals.  We used a template constructed from out-of-transit spectra for the purposes of measuring the RM effect, as RM signals come from minute perturbations to individual line shapes.  We found only a slight degradation in RV precision (an increase in error of around 5 c\ms) when using a template constructed using out-of-transit data instead of using one constructed from all of the available TRAPPIST-1 spectra.  Our spectra have negligible signal below wavelengths of around 730\,nm, and the observed signal is dominated by two of the three reddest MAROON-X orders. 

\section{Analysis}
\label{sec:analysis}

\subsection{Stellar and Planetary Parameters}
\label{ssec:params}

There are several stellar and planetary parameters that drive the exact shape and amplitude of the RM curve.  We describe them in this section and justify the priors used in the final MCMC fit.

In the RM curve, there is a well-known degeneracy between the stellar rotational velocity and the obliquity, our parameter of interest. The amplitude of the RM effect, at small values of $R_p/R_\star$ and ignoring limb darkening, goes as \citep{Winn10b}:

\begin{equation}
    K_{\mathrm{RM}} \propto v~\mathrm{sin} (i) \bigg( \frac{R_p}{R_\star} \bigg)^2 \sqrt{1-b^2}
\end{equation}

where $v\,sin\,i$ is the projected rotational velocity of the star and $b$ is the impact parameter of the planet.  The signal amplitude is related in a more complex fashion to $\lambda$; typically, objects with polar or near-polar orbits will cause no RV deviations.  It is thus helpful to constrain TRAPPIST-1's $v\,sin\,i$, though we do note that this degeneracy is broken if the planetary inclination is not $90\degr$, as in that case the obliquity also has an effect on the shape of the RM curve.

While TRAPPIST-1 has been extensively studied, its rotational velocity is uncertain. \cite{Reiners18}, directly measuring the line broadening with CARMENES, found that TRAPPIST-1 had a $v\,sin\,i$ $<2$\,\kms, consistent with it being a slow-rotating star.  A photometric rotation period measurement of 3.3 days from \cite{Luger17} and \citep{Dmitrienko18} pointed to a $v\,sin\,i$ = 1.8\,\kms, in agreement with the CARMENES measurement.  However, \cite{Morris18} concluded that this 3.3 day signal may be an activity timescale and not a rotation period.  We define a conservative upper velocity limit of $v\,sin\,i < 10$\,\kms and explore how that impacts our results in later sections.  Ideally, as none of the TRAPPIST-1 planets have inclinations of exactly 90\degr, we should be able to break the degeneracy between the obliquity and rotational velocity, allowing us to use a more uninformative velocity prior.

Another important parameter to consider is the stellar limb darkening.  As the RM effect is the result of the planet covering up different portions of the star's surface, the variation in stellar brightness across the surface will have some impact on the final results.  We make use of quadratic limb-darkening parameters for TRAPPIST-1 from \cite{Claret12}, using the parameters for a $T_\mathrm{eff}=2600$ K, log$g$=5.0 star in the SDSS $z$ bandpass, which is the listed bandpass which is the most similar to the highest-SNR MAROON-X orders.  

We did not make an effort to estimate or model the surface convective blueshift of TRAPPIST-1.  The surface convective blueshift of a star has been shown to be relevant to RM curve calculations by \cite{Cegla16}.  However, they found that the difference between a model that does and does not account for convective blueshift is $\leq 0.5$\,\ms for stars with $v\,sin\,i$=2\,\kms.  Given that our typical RV error estimates are on the order of 5 \ms, such an effect will not be detectable in our results, so we avoided modelling the convective blueshift, as this would have introduced additional model complexity.

Given the large-amplitude transit time variations in our datasets and our lack of simultaneous photometry, we estimated the actual transit times in our datasets by making use of the times forecasted by \cite{Agol2021}, along with their forecast errors, which are all less than five minutes.  We corrected the times recorded by MAROON-X to the times at the solar system barycenter, using light travel times calculated by \texttt{astropy}, to allow for direct comparison to the forecast times, which are recorded in BJD$_\mathrm{TDB}$.  The observed RM curves (especially for planets where the signal is especially clear, such as b and f) appear to roughly coincide with the \cite{Agol2021} transit time predictions, supporting our choice to make use of the their forecasts in our analysis. 

TRAPPIST-1's multiple transiting planets \citep[with impact parameters varying between 0-0.4;][]{Agol2021} allow us to measure the differential rotation velocity $\alpha$ across the star's surface.  In general, a higher absolute value of $\alpha$ (which varies between -1 and +1) indicates a more latitude-dependent rotation rate, while a value of $\alpha$ close to zero indicates that the stellar rotation is at the same rate across the entire surface of the star.  As the TRAPPIST-1 planets have different impact parameters, it may be possible to measure $\alpha$ because different planets scan different latitudes of the host star.  \cite{Vida17} found an additional $P=2.9$\,day signal in the K2 TRAPPIST-1 light curve that could potentially be related to some form of surface differential rotation, but noted that that would result in an unphysically high surface shear, meaning that that signal is likely related to some other phenomena.  Thus, the potential presence of surface differential rotation on TRAPPIST-1 is mostly unconstrained.  Typically, stars with rapid rotation (and early M dwarfs) tend to have low $\alpha$ values $\alpha<0.1$ \citep[e.g.,][]{Reinhold14, Kuker19}, so we expect to see something similar for this system, though the surface differential rotation of late M dwarfs is relatively poorly studied.  

The planetary periods, radii and inclinations, as well as the stellar radius, are all constrained in \cite{Agol2021}.  While all of these parameters have an impact on the observed obliquity, they have been constrained to the point that allowing them to vary will be unlikely to meaningfully impact our fits (besides making them take much longer).  As a point of reference, the typical radius error of a TRAPPIST-1 planet (and the star itself) is on the order of 1 -- 2\%, which would translate to a roughly 2 -- 4\% change in the RM amplitude.  Given that our typical RV error bars are on the order of 30 -- 50\% of the RM curve amplitude, fitting the radius would not meaningfully impact our results.  The period error (which would affect the transit duration) is also on the order of tenths of a minute, which is substantially less than the other timescales considered in this analysis, such as the errors on the forecast transit times and the exposure lengths.  Finally, perturbing the individual planets' inclinations by their error bars only results in changes in RV on the order of $<1$\,\ms, which is much less than the typical RV errors.  These deviations are even smaller for the outer planets, with more well-constrained inclinations, and for systems with low obliquities.  Thus, we fix these parameters at the medians of the distributions quoted by \cite{Agol2021} in our fits.  The planetary eccentricities are all known to be low and thus we fix them at $e=0$ in our fits.

The RM curve for an object with an obliquity of $\lambda$ is identical to that with an obliquity of $-\lambda$ for an edge-on planet.  However, the TRAPPIST-1 planets are slightly inclined, breaking this degeneracy.  Thus, we allow $\lambda$ to vary in between -180$\degr$ and 180$\degr$ in our fits.

Additionally, we found linear trends in our RV baselines that can't be explained by the known planetary system around TRAPPIST-1.  We observe slopes on the order of 10\,\ms over three hours, which is not in agreement with the relatively small RV amplitudes of the known planets.  We did not notice any long-term or large-amplitude trends in RV over our entire dataset over two years of observation, so these trends appear to be incoherent over long timescales.  They are likely due to short-term stellar activity, and only obscure what we are actually trying to study.  Thus, we simultaneously fit a linear slope term alongside the RM curve fit for each night of transit observations. 

In our final analysis, the parameters allowed to vary were the stellar rotation velocity $v\,sin\,i$, the planetary obliquities $\lambda$, the stellar surface differential rotation $\alpha$, the transit times $T$, and the nuisance linear trend.  We performed fits both assuming the planets all have a shared obliquity and, alternatively, with all of the obliquities calculated separately.  The former model is a reasonable assumption, as it would be a tremendous coincidence for all of the planets to transit the star at the same inclination but have non-coplanar orbits.  The latter model is useful for showing which planets are the most effective at constraining the overall obliquity.  Unfortunately, it is difficult to estimate the obliquities of some of the low-signal planets individually (such as d and e) with this method.  The values used for each parameter, along with the priors used (if relevant), are shown in Table~\ref{tab:priors}.

\begin{table*}[]
\centering
\begin{tabular}{lccc}
\textbf{Parameter}                      & \textbf{Value} & \textbf{Prior}                                                               & \textbf{Source} \\ \hline
\textbf{Eccentricity}                   & 0              & fixed                                                                        &                 \\ 
$\alpha$                                &                & $\mathcal{U}$(-1, 1)                                                          &                 \\ 
\textbf{Obliquity $\lambda$ (degrees)}  &                & $\mathcal{U}$(-180, 180)                                                     &                 \\ 
\textbf{Stellar radius (\Rsun)}         & 0.1234         & fixed                                                                        &                 \\ 
\textbf{v$sin$i (\ms)}                  &                & $\mathcal{U}$(0, 10000)                                                      &                 \\ 
\textbf{$u_{1, LD}$}                    & 0.6542         & fixed                                                                        & b               \\ 
\textbf{$u_{2, LD}$}                    & 0.2834         & fixed                                                                        & b               \\ 
\textbf{Fit offset} (\ms)               &                & $\mathcal{U}$(-1000, 1000) for each transit                                  &                 \\ 
\textbf{Fit slope}  (\ms/day)           &                & $\mathcal{U}$(-1000, 1000) for each transit                                  &                 \\ 
\textbf{Transit Time (BJD - 2,450,000)} &                &                                                                              &                 \\
$T_{b}$                                 &                &  $\mathcal{N}$(9520.849564, 0.000331)                                        & a               \\
$T_{c}$                                 &                &  $\mathcal{N}$(9104.003866, 0.000285),  $\mathcal{N}$(9786.951957, 0.000656) & a               \\
$T_{d}$                                 &                &  $\mathcal{N}$(9783.965213, 0.003294),  $\mathcal{N}$(9788.015726, 0.003325) & a               \\
$T_{e}$                                 &                &  $\mathcal{N}$(9526.829202, 0.001351), $\mathcal{N}$(9782.991182, 0.001571)  & a               \\
$T_{f}$                                 &                &  $\mathcal{N}$(9788.943700, 0.001198)                                        & a               \\
$T_{g}$                                 &                &  $\mathcal{N}$(9184.835364, 0.000714)                                        & a               \\ 
\textbf{Planet Radius (\Re)}            &                &                                                                              &                 \\
$R_{b}$                                 & 1.116          &  fixed                                                                       & a               \\
$R_{c}$                                 & 1.097          &  fixed                                                                       & a               \\
$R_{d}$                                 & 0.788          &  fixed                                                                       & a               \\
$R_{e}$                                 & 0.920          &  fixed                                                                       & a               \\
$R_{f}$                                 & 1.045          &  fixed                                                                       & a               \\
$R_{g}$                                 & 1.129          &  fixed                                                                       & a               \\ 
\textbf{Planet Inclination (\degr)}     &                &                                                                              &                 \\
$i_{b}$                                 & 89.728         &  fixed                                                                       & a               \\
$i_{c}$                                 & 89.778         &  fixed                                                                       & a               \\
$i_{d}$                                 & 89.896         &  fixed                                                                       & a               \\
$i_{e}$                                 & 89.793         &  fixed                                                                       & a               \\
$i_{f}$                                 & 89.740         &  fixed                                                                       & a               \\
$i_{g}$                                 & 89.742         &  fixed                                                                       & a               \\ 
\textbf{Planet Period (d)}              &                &                                                                              &                 \\
$P_{b}$                                 & 1.510826       &  fixed                                                                       & a               \\
$P_{c}$                                 & 2.421937       &  fixed                                                                       & a               \\
$P_{d}$                                 & 4.049219       &  fixed                                                                       & a               \\
$P_{e}$                                 & 6.101013       &  fixed                                                                       & a               \\
$P_{f}$                                 & 9.207540       &  fixed                                                                       & a               \\
$P_{g}$                                 & 12.352446      &  fixed                                                                       & a               \\ 
\textbf{Semi-Major Axis ($10^{-2}$\,AU)}&                &                                                                              &                 \\
$a_{b}$                                 & 1.154          &  fixed                                                                       & a               \\
$a_{c}$                                 & 1.580          &  fixed                                                                       & a               \\
$a_{d}$                                 & 2.227          &  fixed                                                                       & a               \\
$a_{e}$                                 & 2.925          &  fixed                                                                       & a               \\
$a_{f}$                                 & 3.849          &  fixed                                                                       & a               \\
$a_{g}$                                 & 4.683          &  fixed                                                                       & a               \\ 
\hline
\end{tabular}
\caption{The parameters used in the MCMC fit of the RM effect of the TRAPPIST-1 planets.  Reference a) refers to \cite{Agol2021} and b) refers to \cite{Claret12}.}
\label{tab:priors}
\end{table*}

\subsection{RM Modeling with \texttt{starry}}
\label{ssec:rm}

We modeled the RM effect using the python code \texttt{starry} \citep{starry}, which simulates a rotating limb-darkened stellar surface and estimates the observed RV shifts as a function of transiting planet position.  Adopting the priors from Section~\ref{ssec:params}, we then fit for the $v\,sin\,i$, $\lambda$, $\alpha$, and $T$ using the MCMC sampler \texttt{emcee} \citep{Foreman-Mackey13}.  As our exposures were long compared to the timescale of the transit, we modeled the expected RV of each exposure by calculating the RV at four times within each exposure and then averaging out the values.  This methodology is valid because the RVs evolve smoothly at all times except the very beginning and the very end of the transit.

Initially, we considered two separate models: one in which we assumed all of the planets had the same obliquity (see Table~\ref{tab:share_obl_no_alpha}), and one in which we fit the individual planet obliquities separately (see the table in Appendix~\ref{appendix:fits}).  The model in which all of the obliquities are fixed to the same value are preferred by the BIC ($\Delta$BIC = -16.2), indicating that we lack statistical evidence to conclude that TRAPPIST-1's planets don't share a common obliquity.  This is an expected result, as it would be unlikely for these planets to all have different orbital orientations but transit the host star at the same inclination.  For many of the lower-amplitude and short-period planets (such as c, d, and e), the observed posterior distributions of $\lambda$ were multimodal, reflecting the fact that the RM curve of a planet with $\lambda=x\degr$ looks extremely similar to that of a planet with $\lambda=-x\degr$ when the impact parameter is low.  The planets with the longest orbital periods, TRAPPIST-1f and TRAPPIST-1g, had the most constraining individual transits, with roughly $\pm 20\degr$ constraints on their obliquities from a single transit.  Future efforts at constraining the TRAPPIST-1 system should likely focus on these planets.

\begin{table*}[]
\centering
\begin{tabular}{ccc}
\textbf{Parameter}       & \textbf{Planet} & \textbf{Value}                                           \\ \hline
$v\,sin\,i$ (\kms)               & -               & $2.10 \pm 0.29$                                       \\
$\lambda$ (\degr)                  & -               & $-2^{+17}_{-19}$                                    \\
$T$ (BJD - 2,450,000) &                 &                                                          \\
                         & b               & 9520.8496$\pm 0.0003$                              \\
                         & c               & 9104.0038$\pm 0.0003$, 9786.9523$\pm 0.0007$ \\
                         & d               & 9783.9642$\pm 0.0026$, 9788.0135$\pm 0.0021$   \\
                         & e               & 9526.8290$\pm 0.0013$, 9782.9909$\pm 0.0015$   \\
                         & f               & 9788.9437$\pm 0.0011$                               \\
                         & g               & 9184.8350$\pm 0.0007$                              
\end{tabular}
\caption{The MCMC best-fit parameters of the RM model for an $\alpha=0$ star and a system of planets that all share the same obliquity.}
\label{tab:share_obl_no_alpha}
\end{table*}

In our shared-obliquity model (plotted in Figure~\ref{fig:rm_rv}), we find that the obliquity of the system is equal to $\lambda= -2\degr^{+17\degr}_{-19\degr}$ from the RM effect alone, which is consistent with the planets having no spin-orbit misalignment.  This is an improvement over the RM fits performed on the three planet data set studied by \cite{Hirano2020}, which had a precision of $\pm 28\degr$, but still doesn't allow us to conclude whether or not the TRAPPIST-1 system has a tidally damped obliquity.  We also measure $v\,sin\,i$ = 2.1 $\pm$ 0.3\,\kms, which would correspond to a rotation period (assuming $sin\,i \approx 1$) of about 2.5 -- 3.3 days.  This is consistent within two sigma with the velocity of 1.5 $\pm$ 0.4\,\kms from \cite{Hirano2020} and also agrees with the rotation period of 3.3 days from \cite{Luger17} and \cite{Dmitrienko18}.  The \cite{Agol2021} transit time forecasts describe our results well, though we do note that the g transit appears to occur slightly earlier than the forecast predicts.  

\begin{figure}
    \centering
    \includegraphics[width=1\linewidth]{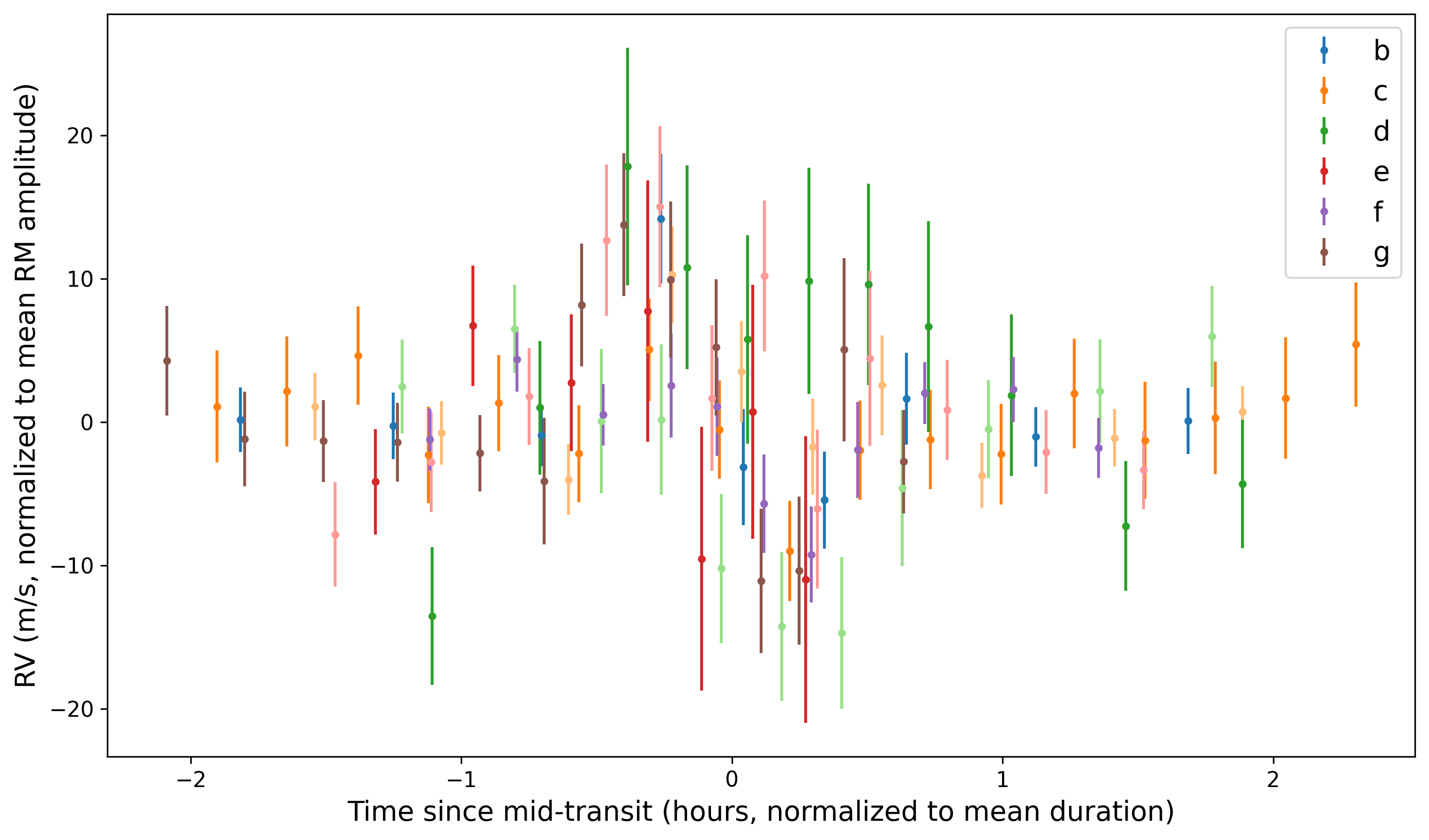}
    \caption{The observed RVs of TRAPPIST-1, stacked and stretched in time and RV amplitude according to the best-fit model to the planets with a shared obliquity and zero $\alpha$.  The planets are each plotted with different colors to highlight which systems were the most and least constraining with respect to the RM fit.}
    \label{fig:rm_stacked}
\end{figure}

Figure~\ref{fig:rm_stacked} highlights the observed RM signal by shifting, stacking, and stretching the dataset according to the best-fit model of each individual planet's transit duration and signal amplitude.  It is obvious that the smallest planets (d and e) had substantial RV errors relative to the expected RM signal and thus were not very helpful in constraining our results.

\begin{figure}
    \centering
    \includegraphics[width=1\linewidth]{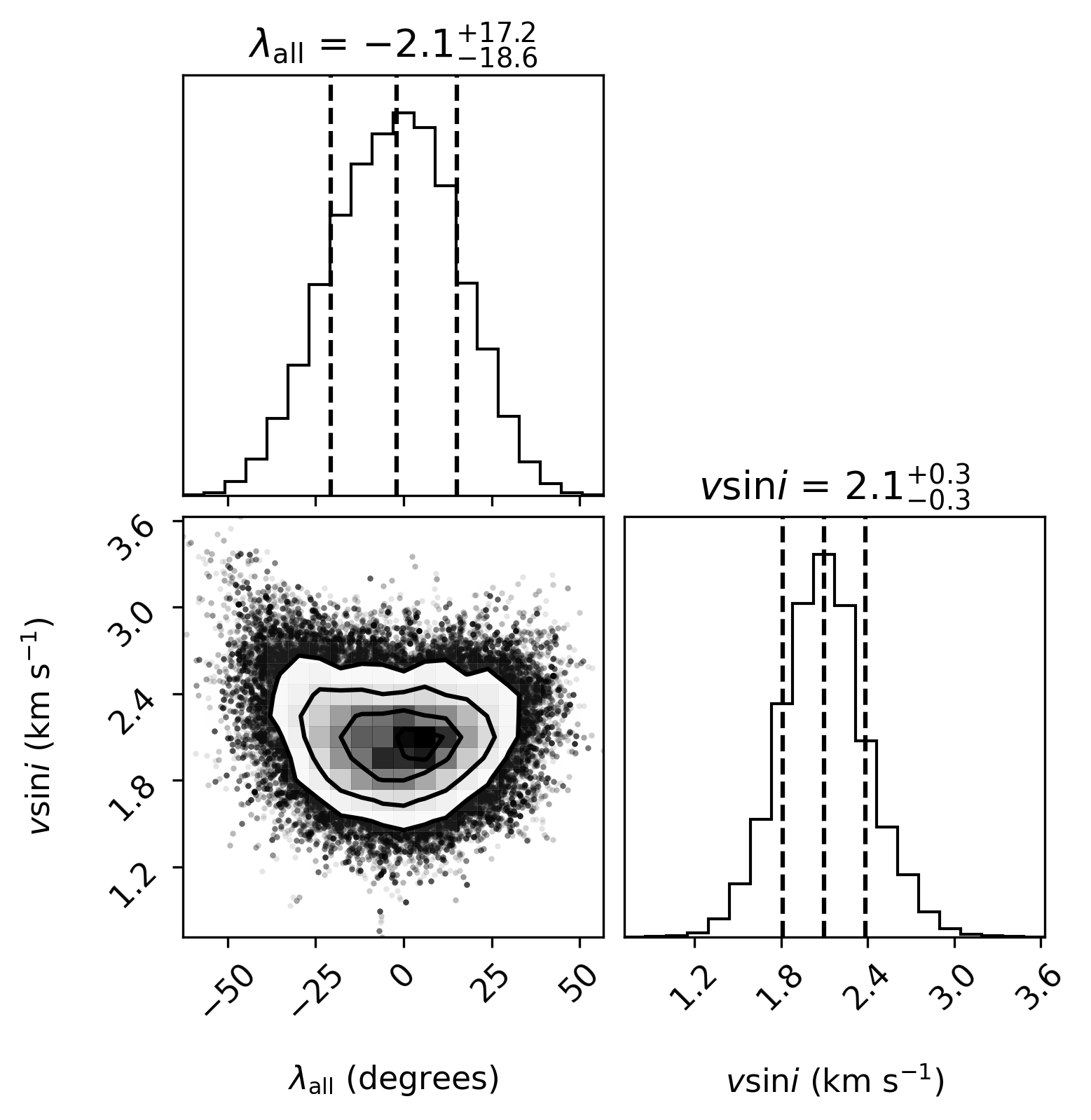}
    \caption{A corner plot of $v\,sin\,i$ and $\lambda$ for our preferred TRAPPIST-1 model (the planets share their obliquity and $\alpha=0$).  The transit times and linear trends have no visible correlations and are thus not shown.}
    \label{fig:rm_bestmodel_corner}
\end{figure}

Figure~\ref{fig:rm_bestmodel_corner} shows a corner plot of this model.  There is a slight degeneracy between $v\,sin\,i$ and $\lambda$, but this only becomes a problem at large values of $\lambda$, which are inconsistent with the data. 

We also investigated the impact of $\alpha$ on our fits (shown in Appendix~\ref{appendix:fits}).  Given the degeneracies between $v\,sin\,i$ and $\lambda$, we only fit for alpha in our model where all of the planets have a shared $\lambda$.  We found that the posterior distribution of $\alpha$ ($\alpha = -0.1^{+0.7}_{-0.6}$) is essentially just the priors (with only a slight preference for low values of $\alpha$), and allowing $\alpha$ to vary causes a slight increase in the BIC over a model in which it is simply fixed at zero, $\Delta$BIC = 2.7. In addition, the fit values for $v\,sin\,i$ (2.1 $\pm$ 0.3\,\kms) and $\lambda$ ($-2 \pm 18\degr$) are very similar to those fit in the case that $\alpha$ is fixed at 0.  Future efforts to measure $\alpha$ will likely require far more precise data or a system featuring planets with a broader range of impact parameters.

\subsection{Estimating TRAPPIST-1's Rotational Velocity}
\label{ssec:vsini}

As we have access to high-resolution spectra from MAROON-X, we can estimate the star's rotational broadening directly from the spectrum by examining its lines.  We note that MAROON-X has a resolution of roughly 85,000 \citep{Seifahrt18}, meaning that we cannot expect to accurately measure rotational velocities below about 2 -- 3\,\kms. In this section, we examine the MAROON-X spectrum to verify that the $v\,sin\,i$ is at or below this limit, which would support the low rotational velocity measured in Section~\ref{ssec:rm}.

To estimate the rotational velocity, we used the cross correlation function (CCF) comparison method described in \cite{Gray05} and employed in \cite{Reiners12}.  Instead of comparing our TRAPPIST spectra directly to rotationally-broadened model spectra (which may introduce systematic biases due to differences between the template and stellar spectra), we calibrated the relationship by studying an as-observed MAROON-X spectrum of a known slow rotator.

First, we selected a spectrum of a star that has been observed by MAROON-X that is similar in spectral type to TRAPPIST-1 but is known to be a slow rotator.  We selected several different stars as templates, as we had no stars that had the same spectral type as TRAPPIST-1 \cite[M7.5,][]{Gizis00} and thus wanted to investigate the impact of varying spectral type on our final results.  The template stars are listed in Table~\ref{tab:vsini_templates}, but all of them have rotation periods suggesting a low $v\,sin\,i \leq 0.1$\,\kms.

\begin{table*}[]
\centering
\begin{tabular}{ccccc}
\textbf{Star}    & \textbf{Type} & \textbf{P$_\mathrm{rot}$ (days)} & \textbf{Period Reference} & \textbf{TRAPPIST-1 $v\,sin\,i$ (\kms)} \\ \hline
Teegarden's Star & M7                     & $99.6 \pm 1.4$                  & \cite{Terrien22}         & $1.9 \pm 0.7$                      \\
LP 791-18        & M6                     & >100                            & \cite{Crossfield19}      & $2.7 \pm 0.7$                      \\
Ross 128         & M4                     & >100                            & \cite{Bonfils18}         & $2.4 \pm 0.5$                      \\
Barnard's Star   & M4                     & $145 \pm 15$                    & \cite{Terrien22}         & $2.4 \pm 0.8$                     
\end{tabular}
\caption{The various stars used as templates for the purposes of estimating the $v\,sin\,i$ of TRAPPIST-1.  The last column shows the derived $v\,sin\,i$ of TRAPPIST-1, using the given star as a template.}
\label{tab:vsini_templates}
\end{table*}

We simulated rotational broadening of the template spectrum for a grid of different values of $v\,sin\,i$ (0 -- 10\,\kms, with a grid spacing of 0.5\,\kms) using the rotational convolution kernel from \citep{Gray08}.  We assumed a linear limb-darkening coefficient of 0.8446 \citep{Claret12}, corresponding to the linear limb-darkening coefficient of a 2600\,K, log$g$=5.0 star in the SDSS \textit{z} filter.  This likely represents a slight overestimate of the limb-darkening coefficient, as most of the template stars are hotter and bluer and thus have slightly lower linear limb-darkening coefficients.  However, we found that decreasing the limb-darkening coefficient by as much as 0.5 tended to only have a minor effect on the resulting calculated rotational velocities, decreasing them by around 0.1 km/s, which is significantly less than the quoted $v\,sin\,i$ error and thus doesn't warrant a more precise prescription.

\begin{figure*}
    \centering
    \includegraphics[width=0.9\linewidth]{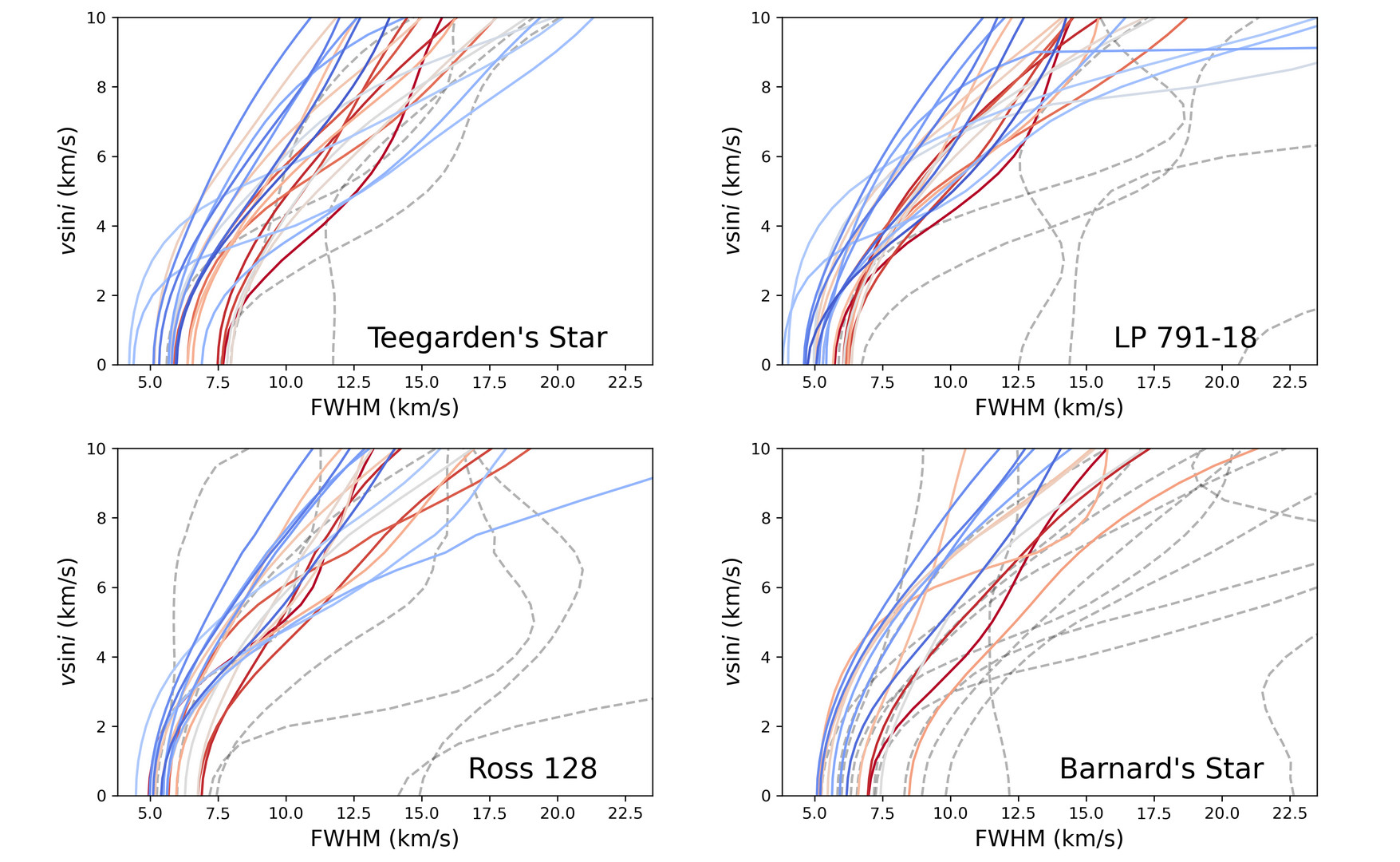}
    \caption{The CCF FWHM as a function of the $v\,sin\,i$ used to broaden the template spectrum.  Each line represents a different order, the the colors determined by the central wavelength of the order (redder orders are redder, bluer orders are bluer).  The dashed gray lines represent orders that weren't used in our final analysis.}\label{fig:fwhm_vsini}
\end{figure*}

We masked the tellurics out of the template's spectrum and calculated the CCFs between the broadened template and the original unbroadened template spectrum for each given rotational velocity.  We then fit the center of each CCF with a gaussian and determined the CCF full width half-maximum (FWHM).  The FWHM of the CCF is thus directly correlated with the template $v\,sin\,i$.  This relationship was fit with a simple quadratic interpolator to create a function that can estimate the $v\,sin\,i$ of an unbroadened stellar spectrum given its CCF FWHM.  This analysis is done on an order-by-order basis, such that each template MAROON-X order has an associated function, shown in Figure~\ref{fig:fwhm_vsini}.  We dropped any orders with non-monotonic functions between $v\,sin\,i$ and the FWHM, as that is an indication that the CCFs in these orders are dominated by systematics that confuse the gaussian fitting process.  This frequently happens in orders with low signals or many masked telluric regions.

As shown in Figure~\ref{fig:fwhm_vsini}, we found that each star has a slightly different relationship between FWHM and $v\,sin\,i$, even when compared to other stars of the same spectral type.  As an example, LP 791-18 has a much narrower CCF than Teegarden's Star, despite them both being late M dwarfs that rotate slowly.  We note the presence of stellar activity, including flaring, in the MAROON-X data for Teegarden's Star, which could possibly account for its slightly broader CCF, though we selected a spectrum that was not taken during a flare for this analysis.  These differences could also be merely due to the differences in spectral type or telluric absorption between these spectra.  We can also see that, below about $v\,sin\,i$ = 2\,\kms, the relationship between FWHM and $v\,sin\,i$ for most of the template stars approaches a vertical line, meaning that this method is incapable of discriminating between $v\,sin\,i$ values significantly below 2\,\kms.  This is an expected result given the resolution of MAROON-X.

We then calculated the CCF of the observed TRAPPIST-1 spectrum with the unbroadened template, and estimated the $v\,sin\,i$ for each order, assuming that the CCF broadening comes primarily from the rotation of TRAPPIST-1.  In some orders, the FWHM of the TRAPPIST-1-to-template CCF was smaller than the FWHM of the unbroadened template CCF referenced against itself, which is an unphysical result that yields a negative $v\,sin\,i$ calculation.  This issue is likely due to systematic differences between the TRAPPIST-1 spectrum and the spectra of the template stars, which could result in slight deviations in the derived line FWHMs.  It could also be a side-effect of a particularly noisy (or telluric-ridden) template spectrum that is ill-suited to a CCF calculation.  We thus excluded these orders from our analysis.  We also excluded orders which fit to rotational velocity values higher than 10\,\kms, which likely suffer from similar problems.  We took the weighted mean and standard deviations of the $v\,sin\,i$ values of the remaining orders to estimate the final $v$sin$i$ and its associated errors (rightmost column of Table~\ref{tab:vsini_templates}).  The calculated $v\,sin\,i$ values are shown in the rightmost column of Figure~\ref{fig:fwhm_vsini}.

The $v\,sin\,i$ derivations for TRAPPIST-1 with each template are all shown in Table~\ref{tab:vsini_templates}.  The quoted values and errors are the weighted mean and standard deviation of the order $v\,sin\,i$ values for each template.  These errors represent underestimates of our true error, which is likely somewhat larger due to systematic differences between TRAPPIST-1 and the template spectra, which may differ in terms of activity levels, metallicity, and spectral type.  These differences between the template spectra and TRAPPIST-1 likely also broaden the calculated CCFs, meaning that these rotational velocity values are likely overestimates.  This also assumes that the template stars have no rotation.  As these stars do rotate, our estimate of TRAPPIST-1's rotation is an underestimate.  However, all of the template stars are expected to have $v\,sin\,i$<0.1\,\kms, so the overall effect this would have on our calibration is an underestimate by approximately $<0.2$\,\kms, which is smaller than our quoted velocity errors.

All of the values fall in the roughly 2 -- 3\,\kms range, which is on the edge of what is distinguishable with MAROON-X.  For our final result, we quote the $v\,sin\,i$ estimated using the Teegarden's Star template, which is both the result with the lowest final $v\,sin\,i$ value and the one with the latest-type spectral template.  Thus, with the CCF method, we find that the rotational velocity of TRAPPIST-1 is around $1.9 \pm 0.7$\,\kms (corresponding to a rotation period of 2.3 -- 5.0 days given $sin\,i\approx 1$), which agrees with the $2.1 \pm 0.3$\,\kms measurement performed in Section~\ref{ssec:rm}.

\subsection{Doppler Tomography}

Another method we can use to measure the obliquity is the Doppler Tomography (also known as the Doppler Shadow) technique \citep[e.g.,][]{Cameron10}, in which the obliquity is inferred from line shape perturbations caused by the planet passing over the rotating star's surface.  As the planet eclipses the rotating stellar disk, the planet distorts the stellar line shape, which often manifests as a "bump'' which moves across the stellar line profile.  If the planet is measured with sufficient time coverage, we can watch how this bump moves in time and velocity space and infer the system obliquity.  As the Doppler Shadow technique and the Rossiter-McLaughlin effect are both observable with high-precision spectra, both can be performed independently using the same dataset to constrain a given star's obliquity.

We attempted to perform a Doppler Tomography analysis with the MAROON-X data.  The overall line profile is calculated by estimating the CCF (cross-correlation function) of the given spectrum with a template or mask.  \texttt{serval} uses a least-squares fitting technique instead of the cross-correlation method, which necessitated that we use a different software for this analysis stage.  We adapted the publicly-available \texttt{raccoon} code \citep{Lafarga20} (which was originally produced with the intention to perform CCF analyses of CARMENES data) to be able to process MAROON-X spectra.  We also adapted \texttt{raccoon} to generate a mask template with $\approx 1000$ lines out of the coadded MAROON-X TRAPPIST-1 spectrum.  We found using the publicly available template for Teegarden's Star (which is a M7 dwarf) did little to change our results, even though the template contained many more ($\approx 5,000$) lines.  Using our template, we calculated the line profile for each TRAPPIST-1 exposure after deblazing each spectrum.  We ignored the contributions to the CCF from orders blueward of around 730\,nm and from orders with significant telluric contamination, where the SNR in the MAROON-X spectra was low enough that the data introduced significant noise in the calculated CCF.  We performed this calculation for each of our MAROON-X spectra to find the average line profile of each observation. 

The CCF profile was binned in increments of 3.5\,\kms, to match the expected velocity resolution of MAROON-X. Not binning the CCF profile in velocity space would result in excess correlations between the line profile values.  However, this does mean that the expected Doppler shadow of TRAPPIST-1 (which is expected to travel from -2 -- +2\,\kms) will only be encompassed by around two independent points in each CCF profile.  

We normalized each CCF to unity by fitting a linear term to each line profile's baseline ($>10$\,\kms from the line center) and then dividing each profile by that fit line.  For each night, we calculated an average out-of-transit profile by averaging together all of the line profiles from observations not taken during the transit \citep[according to the predictions of][]{Agol2021}.  We then estimated the residuals by subtracting the night's out-of-transit line profile from each individual CCF profile.  For a well-aligned low-obliquity system, the residuals are expected to show a small bump, traveling from the left side of the line profile to the right side of the line profile, as the planet transits in front of its host. 

To model the signal, we used the models described in \cite{Cameron10}, who modeled the effect of the planetary signal as a gaussian perturbation added to a gaussian line profile with linear limb-darkening.  In this model, we adopted the same parameters for the planetary properties as described in Table~\ref{tab:priors}, but instead adopted a linear limb darkening term of 0.8446 \citep{Claret12} and fit for the $v\,sin\,i$, CCF FWHM, and (shared) planetary obliquity using the methodology described in \cite{Cameron10}, with fits performed using \texttt{emcee}. The FWHM describes the width of the gaussian residual bump in the CCF profile as the planet transits, the $v\,sin\,i$ describes the extent of the bump's motion in velocity space, and the $\lambda$ describes the actual direction that the bump moves in. Due to the low time resolution of our data, much like in Section~\ref{ssec:rm}, we calculated the model CCF at four evenly spaced times encompassed by each observation and averaged them together.  To speed up our fits, we calculated a set of line profiles \citep[referred to as $h(x)$ in][]{Cameron10} based on a grid of planet positions and stellar rotational velocities before running the MCMC fit and performed linear grid interpolation to estimate the line profile at each step.  This dramatically sped up the fit by avoiding the repeated numerical integration that is necessary in their methods, in return for a slight loss in accuracy (typically on the order of one part in $10^5$ or less).

We assumed a fixed relative system velocity of 0\,\kms, as the CCF is generated by referencing the spectra against a template generated from the same spectral data.  The spectra are shifted according to the expected barycentric velocity each night, but we did not include any planet-induced RV shifts because they are substantially smaller than the 3.5 km/s resolution.  We used the same obliquity prior given in Table~\ref{tab:priors}, but adopted an upper limit on the $v\,sin\,i$ as that of the line profile FWHM, as the FWHM is by definition equal to or larger than the $v\,sin\,i$ in the \cite{Cameron10} models.  

\begin{figure*}
    \centering
    \includegraphics[width=0.9\linewidth]{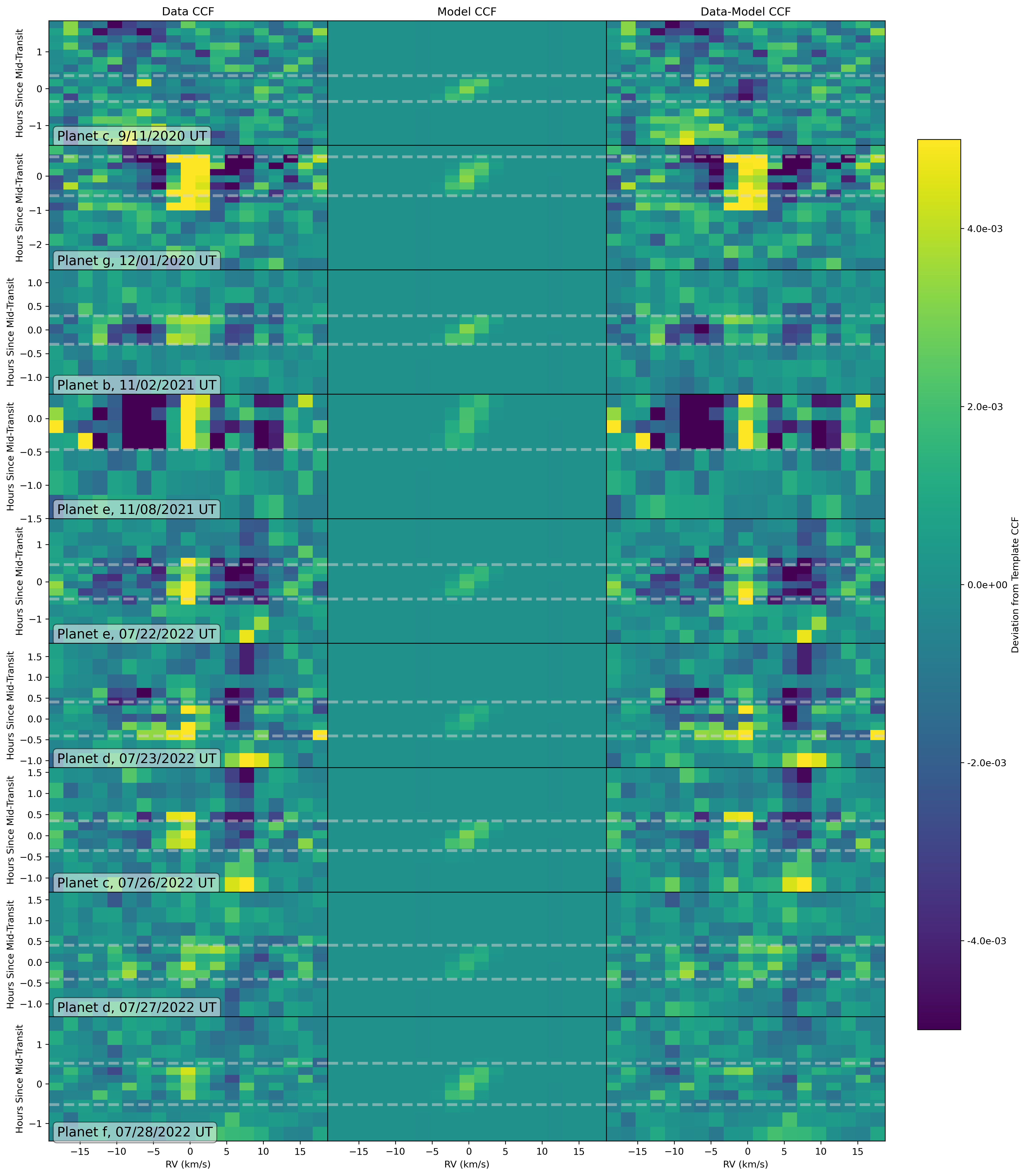}
    \caption{The Doppler tomography signals of the nine observed transits, in order of observation date (c, g, b, e, e, d, c, d , and f).   The leftmost column shows the data residuals without accounting for the Doppler Shadow. The center column shows the expected tomographic signal for a Doppler tomography model, plotted on the same color axis as the data.  The model has a $v\,sin\,i$=2.1\,\kms and $\lambda=0\degr$, in accordance with what was found in Section~\ref{ssec:rm}, and a FWHM=3.2\,\kms, which is a rough estimate of the FWHM taken directly from the shape of the out-of-transit CCF profile.  The right column shows the data residuals with the model subtracted.  The gray dashed lines show the beginning and end times of each transit. The data in-transit appear noisier than the data out-of-transit primarily due to the shorter exposure times.  The velocities are binned in increments of 2\,\kms, which is slightly lower than the binning of 3.5\,\kms used in our analysis.  This is primarily for the purposes of demonstrating the theoretical shape of the modeled signal, which is difficult to display with a coarser binning.}
    \label{fig:doppler_tomography}
\end{figure*}

After processing the data, we compared our simulations to the models described in \cite{Cameron10}.  Overall, we found that the expected signal amplitude of the TRAPPIST-1 planets is comparable to (or less than) the observed noise in the nightly CCFs outside of the line profile (where we don't expect to see any planetary signals).  The data (see Figure~\ref{fig:doppler_tomography}) show a trend of increased noise in the in-transit line profiles and a general tendency for the in-transit line-profiles to be shallower than the out-of-transit line profiles.  This could be explained with planetary signals but could also be the result of slight normalization errors, which are difficult to correct given the small number of data points and the significant noise.  We found that allowing the $v\,sin\,i$ to vary freely resulted in fits that strongly preferred extremely low values of $v\,sin\,i$ ($\approx 1$\,\kms), in disagreement with our results from the canonical RM effect modeling in Section~\ref{ssec:rm} and consistent with a velocity of zero given our instrumental resolution.  Fixing $v\,sin\,i$ at 2.1\,\kms typically resulted in fit obliquity measurements of around 60$\degr$ -- 90$\degr$.  This apparent degeneracy between $v\,sin\,i$ and $\lambda$ is a sign that we have failed to detect the signature of TRAPPIST-1, as a low $v\,sin\,i$ and a $\lambda$ close to $90\degr$ both manifest as a bump traveling up the center of the CCF profile, which can easily be reproduced by slight errors in CCF normalization that are difficult to correct for with our current dataset.

This analysis shows that we are unable to detect the Doppler shadow of TRAPPIST-1 with MAROON-X.  This is primarily due to a combination of the low SNR of the TRAPPIST-1 spectrum, the insufficient resolution of the instrument, and the small expected signal amplitude.  It is obvious from Figure~\ref{fig:doppler_tomography} that the typical CCF noise can be far stronger than the anticipated signal, and our instrument's resolution is too low for us to clearly resolve the planet-induced Doppler shadow.  Additionally, the center of the CCF profile is sensitive to errors in normalization (primarily introduced by our low resolution), making it difficult to distinguish a Doppler tomographic signal from noise unless it is fairly strong.  

\cite{Hirano2020} claimed a detection of this signal based on Subaru IRD data, measuring an obliquity of $19\degr^{+13\degr}_{-15\degr}$.  We note that the IRD is sensitive to redder wavelengths than MAROON-X, which theoretically would make late M dwarf profile characterization more straightforward due to the decreased amount of line blending \citep{Onehag12}.  This would make the generation of an accurate CCF template easier.  In addition, while their derived RVs are lower in precision than those obtained via \texttt{serval}, the MAROON-X RVs derived via CCF (using \texttt{raccoon}) have much larger errors, likely due to the relatively small number of lines included in the template.  They also have shorter exposure times (300 seconds versus 600 seconds) during the transits, which translate to better temporal resolution of the signal.  However, the slow rotational velocity of the star ($\approx 2\,\kms$ and the velocity resolution of IRD ($\approx 4\,\kms$) would likely result in similar problems with regards to fitting the signal as to what we faced with the MAROON-X dataset.  They also found planet signals which were similar in magnitude, if not weaker, than the typical noise in their CCF profiles.  These issues are reflected somewhat by the high false alarm probability of $>1\%$ quoted in the paper.  Overall, we recommend using obliquity and velocity measurements derived from the RM effect instead of those derived from Doppler tomography for this system until a less ambiguous detection is made.

\section{Conclusions}
\label{sec:results}

Using RM measurements, we show that the TRAPPIST-1 planets possess a low obliquity of $-2\degr^{+17\degr}_{-19\degr}$ and a slow rotation velocity of $2.1 \pm 0.3$\,\kms, agreeing with observations from \citep{Hirano2020} and observations of planets around other M dwarfs \citep{Hirano20b, Stefansson20}.  Given the fairly large $a/R_\star$ values of the TRAPPIST-1 planets, it is unlikely that the system has undergone significant obliquity damping over the course of its lifetime, meaning that it is probable that the observed planetary system formed at a low obliquity and is currently not being torqued by some external companion.   

This study also demonstrates MAROON-X's ability to spectroscopically characterize faint systems.  While TRAPPIST-1 is an extremely difficult M dwarf to study (due to its slow rotation, faintness, and short planetary transits), we are still capable of deriving a $\pm 18\degr$ obliquity constraint (under the assumption that all of the planets share the same obliquity) with RM observations alone.  This is a significant improvement over the RM constraints derived by \cite{Hirano2020}, which found a $\pm 28\degr$ constraint (with similar assumptions).  This result highlights the potential for MAROON-X to constrain obliquities around nearby M-dwarf systems which have previously been difficult to characterize, allowing for further studies into how tides affect the orbits of planets around small stars.

We also perform a direct measurement of the rotational velocity from the line profile broadening, finding an estimated $v\,sin\,i$ of $1.9 \pm 0.7$\,\kms.  This value has a much higher error than that derived via the RM effect due to the combined effects of the relatively low signal in the bluer TRAPPIST-1 orders, systematic differences between TRAPPIST-1 and other late-type M dwarfs observed with MAROON-X, and the resolution of MAROON-X.  However, it does agree closely with the value found via the RM effect and shows that TRAPPIST-1 is a slow rotator.

With MAROON-X, we also attempted to measure the Doppler tomographic signal of TRAPPIST-1.  The expected signal was comparable to the observed noise in the line profile residuals, and we were unable to measure a signal that was in agreement with our results from the RM effect.  This is not unexpected, given the relatively low-SNR spectrum and the weak anticipated signals.  The $\approx$3.5\,\kms velocity resolution of MAROON-X also hampers our ability to search for the signal, which is expected to occur on similar (or smaller) velocity scales.  It is thus extremely easy for noise to masquerade as a planetary signal in our MAROON-X data, as our results are very sensitive to how the data are normalized.  As the signal is small and the MAROON-X instrumental resolution is similar to (if not slightly better than) than what was found with IRD, it seems like the detected tomographic signal in \cite{Hirano2020} may have also been a false alarm, though their shorter exposure times and redder wavelength coverage may have marginally improved their ability to resolve the orbit of TRAPPIST-1.  Overall, this is an extremely challenging target for Doppler tomography due to the combination of its high magnitude, small planets, and slow rotational velocity, and we recommend using the RM effect obliquity measurements until we can find a way to reliably minimize these problems.

\vskip 5.8mm plus 1mm minus 1mm
\vskip1sp

This material is based upon work supported by the National Science Foundation Graduate Research Fellowship under Grant No.\ DGE 1746045. The University of Chicago group acknowledges funding for the MAROON-X project from the David and Lucile Packard Foundation, the Heising-Simons Foundation, the Gordon and Betty Moore Foundation, the Gemini Observatory, the NSF (award number 2108465), and NASA (grant number 80NSSC22K0117). We thank the staff of the Gemini Observatory for their assistance with the commissioning and operation of the instrument. This work was enabled by observations made from the Gemini North telescope, located within the Maunakea Science Reserve and adjacent to the summit of Maunakea. We are grateful for the privilege of observing the Universe from a place that is unique in both its astronomical quality and its cultural significance. GS acknowledges support provided by NASA through the NASA Hubble Fellowship grant HST-HF2-51519.001-A awarded by the Space Telescope Science Institute, which is operated by the Association of Universities for Research in Astronomy, Inc., for NASA, under contract NAS5-26555. GS acknowledges support through the Henry Norris Russell Fellowship at Princeton during the preparation of this manuscript.  We thank Eric Agol for his assistance with interpreting the forecast transit times.  We also thank Lars A. Buchhave, Nestor Espinoza, and Neale Gibson for providing a contemporary TRAPPIST-1c transit time obtained from JWST observations under program GO 2440 (PI: Alexander Rathcke) for use as a sanity-check.  This research has made use of NASA's Astrophysics Data System Bibliographic Services.

\software{\texttt{astroplan} \citep{Astroplan}, \texttt{astropy} \citep{Astropy1, Astropy2}, \texttt{barycorrpy} \citep{Kanodia_2018}, \texttt{corner}  \citep{corner}, \texttt{emcee} \citep{Foreman-Mackey13}, \texttt{h5py} \citep{collette_python_hdf5_2014}, \texttt{numpy} \citep{numpy}, \texttt{pandas} \citep{reback2020pandas, mckinney-proc-scipy-2010}, \texttt{PyAstronomy} \citep{PyAstronomy}, \texttt{python3} \citep{Python3}, \texttt{raccoon} \citep{Lafarga20}, \texttt{scipy} \citep{2020SciPy-NMeth}, \texttt{serval} \citep{Zechmeister20}, \texttt{spectres} \citep{Carnall17}, \texttt{starry} \citep{starry}.}

\facilities{Gemini-North (MAROON-X)}

The TRAPPIST-1 transit observations were collected under the programs GN-2020B-Q-115, GN-2021B-Q-122, and GN-2022A-Q-119.  The Teegarden's Star and Ross 128 spectra were also collected under GN-2022A-Q-119.  The Barnard's Star spectrum was collected under GN-2022A-LP-202, and the LP 791-18 observation was collected under GN-2022A-Q-120.

\bibliography{manuscript}

\appendix
\section{Non-selected Model Fit Parameters}
\label{appendix:fits}

\begin{table*}[h]
\centering
\begin{tabular}{ccc}
\textbf{Parameter}       & \textbf{Planet} & \textbf{Value}                                           \\ \hline
$v\,sin\,i$ (\kms)                & -               & 2.54 $^{+0.42}_{-0.39}$                                           \\
$\lambda$ (\degr)                  & -               &                                                          \\
                         & b               & $8^{+30}_{-32}$                                           \\
                         & c               & $39^{+20}_{-60}$                                         \\
                         & d               & $-24^{+61}_{-37}$                                        \\
                         & e               & $44^{+22}_{-31}$                                         \\
                         & f               & $-46^{+23}_{-17}$                                        \\
                         & g               & $4^{+21}_{-23}$                                         \\
$T$ (BJD - 2,450,000) &                 &                                                          \\
                         & b               & 9520.8495 $\pm 0.0003$                              \\
                         & c               & 9104.0039 $\pm 0.0003$, 9786.9522 $\pm 0.0007$ \\
                         & d               & 9783.9640 $\pm 0.0026$, 9788.0137 $\pm 0.0024$   \\
                         & e               & 9526.8289 $\pm 0.0014$, 9782.9907 $\pm 0.0014$   \\
                         & f               & 9788.9441 $\pm 0.0010$                               \\
                         & g               & 9184.8350 $\pm 0.0007$               
\end{tabular}
\caption{The MCMC best-fit parameters of the RM model for an $\alpha=0$ star and a system of planets that are allowed to have separate obliquities.}
\label{tab:no_share_obl_no_alpha}
\end{table*}

\begin{table*}[h]
\centering
\begin{tabular}{ccc}
\textbf{Parameter}       & \textbf{Planet} & \textbf{Value}                                           \\ \hline
$v\,sin\,i$ (\kms)               & -               & 2.09 $^{+0.31}_{-0.30}$                                           \\
$\lambda$ (\degr)                  & -               & $-2 \pm 18$                                         \\
$\alpha$                 & -               & $-0.09^{+0.70}_{-0.63}$                                   \\
$T$ (BJD - 2,450,000) &                 &                                                          \\
                         & b               & 9520.8496 $\pm 0.0003$                              \\
                         & c               & 9104.0039 $\pm 0.0003$, 9786.9522 $\pm 0.0007$ \\
                         & d               & 9783.9644 $\pm 0.0025$, 9788.0135 $\pm 0.0021$   \\
                         & e               & 9526.8291 $\pm 0.0012$, 9782.9909 $\pm 0.0014$   \\
                         & f               & 9788.9437 $\pm 0.0010$                               \\
                         & g               & 9184.8350 $\pm 0.0007$                  
\end{tabular}
\caption{The MCMC best-fit parameters of the RM model for an $\alpha \neq 0$ star and a system of planets that all share the same obliquity.}
\label{tab:share_obl_alpha}
\end{table*}

\end{document}